\documentclass[submission,copyright,creativecommons]{eptcs}
\usepackage{tabto}
\usepackage{tabularx}
\usepackage{multirow}
\usepackage{amssymb}
\usepackage{amsfonts}
\usepackage{graphicx}
\usepackage{url}
\usepackage{listings,multicol}
\usepackage{color}
\usepackage{subfigure}
\usepackage{paralist}

\usepackage{amsmath}
\providecommand{\event}{ESSS 2014} 

\usepackage{framed} 

\newcounter{defitemcnt}
\newcommand{\defitem}{%
\addtocounter{defitemcnt}{1}%
(\arabic{defitemcnt})}

\newcounter{defexamplecnt}
\newcommand{\defexample}{%
\addtocounter{defexamplecnt}{1}%
\arabic{defexamplecnt}}

\newcommand{\JK}[1]{\textcolor{blue}{}}
\renewcommand{\DH}[1]{\textcolor{red}{}}

\newcommand{\PW}[1]{\mathcal{P}(#1)}

\newcommand{\id}[1]{\begin{ttfamily}#1\end{ttfamily}}
\newcommand{\example}[1]{\textit{\textbf{Example \defexample:}} \textit{#1}}

\definecolor{DarkGray}{rgb}{0.3,0.3,0.3}
\definecolor{LightGray}{rgb}{0.6,0.6,0.6}
\definecolor{Green}{rgb}{0,0.6,0}

\lstdefinelanguage{PseudoCode}{
 keywords={X, var, while, foreach, in, do, done, if, then, else, endif, true, false,
function, endfunc, return, null, X}
 keywordstyle=\color{Green}\textbf,
 ndkeywords={class, export, boolean, throw, implements, import, this},
 ndkeywordstyle=\bfseries,
 identifierstyle=\color{black},
 sensitive=false,
 comment=[l]{//},
 morecomment=[s]{/*}{*/},
 commentstyle=\color{Green}\textit,
 stringstyle=\color{DarkGray}\textit,
 morestring=[b]',
 morestring=[b]"
}

\title{Data-flow Analysis of Programs with Associative Arrays\thanks{This work was partially supported by the Grant Agency of the Czech Republic project 14-11384S and by Charles University Foundation grant 431011 and by Charles University institutional funding SVV-2014-260100.}}
\author{David Hauzar, Jan Kofro\v n, and Pavel Ba\v steck\'y
\institute{Department of Distributed and Dependable Systems\\
Faculty of Mathematics and Physics, Charles University in Prague\\
Czech Republic}
\email{\{hauzar, kofron\}@d3s.mff.cuni.cz, anebril@seznam.cz}
}
\def\titlerunning{Data-flow Analysis of Programs with Associative Arrays}
\def\authorrunning{D. Hauzar, J. Kofro\v n, P.  Ba\v steck\'y}

\begin{document}

\maketitle
\vspace{-9pt}
\begin{abstract}
Dynamic programming languages, such as PHP, JavaScript, and Python, provide built-in data structures including 
associative arrays and objects with similar semantics---object properties can be created at run-time and accessed via arbitrary expressions.
While a high level of security and safety of applications written in these languages
can be of a particular importance (consider a web application storing sensitive data and providing its functionality worldwide),
dynamic data structures pose significant challenges for data-flow analysis making traditional static verification methods both unsound and imprecise.
In this paper, we propose a sound and precise approach for value and points-to analysis of programs with associative arrays-like data structures, upon which data-flow analyses can be built. 
We implemented our approach in a web-application domain---in an analyzer of PHP code.
\end{abstract}

\vspace{-19pt}
\section{Introduction}
\vspace{-6pt}
\DH{Nekde bychom meli vyjasnit, co myslime tim, ze index/property can be accessed via firts-class names. Takto to pouzivali v~\cite{Sridharan:2012}, ale nejsem si jisty, ze to je uplne zrejme. Myslime tim, ze index/property, ke kteremu se pristupuje (na dane urovni), muze byt specifikovany libovolnym read-pristupem k jinemu indexu/property (tento pristup muze zahrnovat libovolny pocet urovni a muze byt specifikovan pomoci staticky neznamych hodnot). Pokud se jenom rekne, ze index/property is accessed dynamically, neni jasne, jestli nemuze byt cteny index/property specifikovany pouze top-level promennou. Takze mozna neco jako: we treat an read-access to associative arrays-like structure of an arbitrary depth as a first class name. Note that the read access can be specified by another read access at an arbitrary level. Ale takhle je to dost slozite. Takze jak presne to napsat? Kam to dat? Mozna tuhle informaci nedat uplne najednou, ale rozprostrit ji---na zacatku by mel ctenar nejakou predstavu, ktera by se postupne zpresnovala.}
Dynamic languages are often used for the development of security-critical applications, in particular web applications.
To assure a reasonable level of reliability, various static analyses such as security analysis and bug finding  are applied. In these cases, data-flow analysis is a necessary prerequisite.
Unfortunately, dynamic features pose major challenges here.
For instance, any interprocedural data-flow analysis needs to track types of variables to determine targets of virtual method calls.
This becomes even more important in the case of languages with dynamic type systems, where types of variables can be completely unspecified.
Moreover, names of methods to be called or files to be included can be computed at run-time. 
Next, all these data can be manipulated using built-in dynamic data structures, such as multi-dimensi\-onal associative arrays and objects with similar semantics---object properties can be created at run-time and accessed via arbitrary expressions, e.g., variables.
This happens relatively often, e.g., in web applications, which manipulate a lot of input.

Tracking values of variables and analysing the shape of these data structures is thus essential for data-flow analysis of dynamic languages.
In this paper we present our approach to these challenges. Our contribution includes: 
(1) value and points-to analysis of associative arrays with an arbitrary depth and accessible using statically unknown values and arbitrary expressions, 
(2) modeling explicit aliases, and (3) a prototype implementation.

%
%



\vspace{-17pt}
\section{Motivation and Overview}
\vspace{-11pt}
In this section, we show some dynamic features that impact the data-flow analysis and present an overview of our approach.
We use PHP
as the representative of a dynamic language; 
However, other languages, especially those connected with the development of web applications, provide
built-in associative arrays-like data structures.
These languages includes Java-script, Python, Ruby, etc.
Moreover, libraries of ``ordinary'' programming languages emulate some of these features and offer the developer API behaving in a similar way.
Hence, our approach is not limited to PHP.
For the illustration of our concepts, we use the code in Fig.~\ref{fig:phpcodeex1} as a running example.

\begin{figure}[ht]
\centering
\begin{framed}
\vspace{-8pt}
\begin{lstlisting}[multicols=2]
$any = $_GET['user_input'];  // an arbitrary user input
$alias = 1; $alias2 = 1; $alias3 = 1;
if ($any) {
	$arr[$any] = &$alias;
	$t = $arr[1]; // t can be either undefined or can have value 1
	$t[2] = 2; // can update also $alias[2] and e.g. $arr[1][2]
	$arr[1][2] = 3;
	$arr[1][3] = 4;
	$arr[2][3] = 5;
} else {
	$arr[$any][2] = 6;
	$arr[1][$any] = 7; // can update also some of variables involved by the previous update
}

$arr[2][1] = &$alias2; // $arr[2][1] and $alias2 can be aliased also with $alias[1]
$arr[2] = &$alias3;
$arr2 = $arr; // deep-copies $arr, including aliases
$arr2[2] = 8; // updates also $arr[2] and $alias3
$arr2[3] = 9; // can update also $arr[3] and $alias
$arr[$any] = arr2;


\end{lstlisting}
\vspace{-5pt}
\end{framed}
\vspace{-1pt}
\caption{Running example.}
\label{fig:phpcodeex1}
\vspace{0pt}
\end{figure}

\vspace{-11pt}
\subsection{Variables, Arrays, and Objects}
\vspace{-3pt}
Variables as well as indices and object properties need not be declared.
If a specified index exists in an array, it is overwritten; if not, it is created.
At line 7 in Fig.~\ref{fig:phpcodeex1}, a new array is created in \id{\$arr} and index \id{2} is added to this array. Next, at line 8, index \id{3} is added to this array.

Arrays 
can have an arbitrary depth.
Unfortunately, updates of such structures cannot be decomposed.
That is, splitting the update at line 7 into two updates at lines 5-6 results in different semantics.
The first reason is that the array assignment statement deep-copies the operand.
The update at line 6 thus does not update the array stored at \id{\$arr[1]}, but its copy.
The second reason is that while updates create indices if they do not exist, read accesses do not;
  while the update at line 7 creates an index containing an array in \id{\$arr[1]} in the case it does not exist, the read access at line 5 returns \id{null} in this case and the update at line 6 fails.

The semantics of the PHP object model is similar to the semantics of associative arrays. 
Objects' properties need not to be declared. If a non-existing property is written, it is created. 
As well as indices, properties can be accessed via arbitrary expressions. 
Objects can also have an arbitrary depth in the sense of reference chains.
In the following, we describe associative arrays, however, the same principles apply to objects as well. 
We write associative arrays-like data structures to emphasize this fact.

\vspace{-10pt}
\subsection{Dynamic Accesses}
\vspace{-3pt}
In dynamic languages, variables, indices of arrays, and properties of objects can be accessed with arbitrary expressions. 
At line 4 in Fig.~\ref{fig:phpcodeex1} the \id{\$arr} array with an index determined by the variable \id{\$any} is assigned;
if a given index exists in \id{\$arr}, it is overwritten; if not, it is created.
Therefore, the set of variables, array indices and object fields is not evident from the code.

An update can involve more than one element and can be statically unknown. 
The update at line 4 is statically unknown and thus may or may not influence accesses at lines 5, 7, 8, 9, 15, and 20.
Similarly, line 11 can access index \id{2} in any index at the first level.
In particular, it can access also index \id{1} at the first level, which is updated at the following line.
That is, reading \id{\$arr[1][2]} can return either of values \id{6}, \id{7}, and \id{undefined}, reading \id{\$arr[1][1]} can return \id{7} and \id{undefined}, reading \id{\$arr[2][2]} can return \id{6} and \id{undefined}, and reading \id{\$arr[2][1]} always returns \id{undefined}.
Next, after two branches of the if statement are merged at line 13, reading of \id{\$arr[1][2]} can return values \id{6}, \id{7}, \id{3}, and \id{undefined}.

\vspace{-9pt}
\subsection{Explicit Aliasing}\label{expaliasing}
\vspace{-3pt}
PHP makes it possible for a variable, index of an array, and property of an object to be an alias of another variable, index, or property.
After an update of an element, all its aliases are also updated.
Aliasing in PHP is thus similar to references in C++ in many aspects. 

Unlike C++, in PHP each variable, index, and property can be aliased and later un-aliased from its previous aliases and become  an alias of a new element.
As an example, the statement at line 16 un-aliases \id{\$arr[2]} from its previous aliases.
Moreover, a variable can be an alias of another variable only at some paths to a given program point, e.g., if it is made an alias in a single branch of an if statement.

The statement at line 4 makes variable \id{\$alias} an alias of a statically unknown index of array \id{\$arr}.
Hence, the statement at line 7 accesses \id{\$arr[1][2]} and may also access \id{\$alias[2]}.
Similarly, the statement at line 15 makes \id{\$alias2} an alias of \id{\$arr[2][1]} and may also make it an alias of \id{\$alias[1]}.
If an array is assigned, it is deep-copied. However, if an index in the source array has aliases, the set of aliases in the corresponding index in a target array consists of these aliases and the source index.
Consequently, the statement at line 18 updates also \id{\$arr[2]} and its alias \id{\$alias3}.
Similarly, the statement at line 19 may update also \id{\$arr[3]} and \id{\$alias}, because the statement at line 3 may make these aliases of each other.

%

\vspace{-8pt}
\subsection{Overview of the Approach}
\vspace{-3pt}
Our approach consists of the following key parts: 
\begin{inparaenum}[(1)]
\item definition of analysis state,
\item definition of read accesses to associative arrays,
\item definition of write accesses to associative arrays (i.e. the transfer function), and
\item definition of merging associative arrays (i.e. the join operator).
\end{inparaenum}

The fundamental part of analysis state consists of represenation of associative arrays.
Each associative array contains a set of indices including a special index called \emph{unknown field}.
This index stores information that has been written to statically unknown indices of the array.
All indices (including unknown fields) can contain a set of values and also can point to another associative array---the next dimension.
Next, unknown fields always contain value \id{undefined}.
Note that (multi-dimensional) unknown fields allow for dealing with statically unknown accesses, however, they pose a challenge both to definition of new indices and definition of merging associative arrays.

Both indices that are read by a read access and indices that are updated by a write access to a multi-dimensional associative array are specified using a list of expressions.
Each expression specifies an access to one dimension of the array.
At each level, indices corresponding to values of an expression corresponding to the level are followed.
If the expression yields a statically unknown value, all indices (including the unknown field) are followed.
The read access differs from the write access in the way it 
handles the case when there is no corresponding index defined.
While the read access follows the unknown field, the write access defines the index.
Note that in the latter case, all data that could have been assigned to the new index using statically unknown updates are copied to this index.
The reason is that these data are stored in unknown fields and unknown fields are not followed by statically known read accesses.
Next, unlike a read access, a write access distinguishes indices that certainly must be updated and indices that only may be updated.
Former indices are strongly updated---original data are replaced with new data, the latter indices are weakly updated---original data are joined with new data.

\example{At line 5, value \id{1} is used to read-access the first dimension of an associative array.
Because the array has no index corresponding to this value, the unknown field is followed.
It contains value \id{1} from the update at line 4 and also value \id{undefined}.}

\example{
At line 12, value \id{1} is used to write-access the first dimension of an array \id{\$arr}.
The array has no index corresponding to this value and the index \id{\$arr[1]} is thus created.
Because statically unknown assignment at line 11 could involve also the index \id{\$arr[1]}, the data from this assignment are copied to a new index and thus also the index \id{\$arr[1][2]} (with values \id{6} and \id{undefined}) is defined.}

The principal challenge of merging multi-dimensional associative arrays with unknown fields is to determine the set of indices of the resulting array.
That is, the resulting array may contain indices that are not present in any array being merged.
The reason again stems from the fact that unknown fields are not followed by statically known read accesses.

\example{
As an example, see the join point at line 13. 
The array \id{\$arr} contains all indices that are defined in either of merged branches plus the index \id{\$arr[2][2]} (with values \id{6} and \id{undefined}).
The reason of creating new index is that while in the second branch the read access to this index in the first level follows the unknown field and then reaches the value assigned at line 11, in the merged array the read access follows the index \id{\$arr[2]}.}

\vspace{-9pt}
\section{Formalization}
\vspace{-6pt}

We formalized our data-flow analysis using data-flow equations for the forward data-flow analysis~\cite{Nielson:1999}.
The formalization includes handling associative arrays of unlimited depth and accesses with arbitrary expressions to such structures.
It does not explicitly include handling of objects.
While objects are treated analogously to arrays in our implementation,
there are subtle differences.
Therefore, we excluded handling of objects from our formalization to make it more clear.

\vspace{-10pt}
\subsection{Analysis State Space}
\vspace{-4pt}
Tab.~\ref{tab:statespace} presents elements of the state space of our data-flow analysis.
Every state contains a 
variable, which represents the symbol-table.
Because top-level variables can be accessed dynamically (\id{\$\$var} is a variable whose name is given by a value of variable \id{\$var}), we model top-level variables as indices of the symbol-table variable\footnote{Consequently, the notions of index and variable refer to the same abstraction and we use them interchangeably.
That is, an index of an associative array in an arbitrary depth is a first class name the same way as a variable.}.
Function $Map$ maps a variable to a set of its possible values.
Function $\textit{Index}$ maps a variable and an index name to a variable containing an array which has the first variable on the index with this name.
In the following, we say that variable $v$ is an index of variable $p$ identified by value $ind$ if $((v, ind), p) \in \textit{indexOf}$ (i.e., $v$ is {$p["ind"]$}).
A pair of relations $Aliases$ relates variables that are \emph{must} and \emph{may} aliases.
Tab.~\ref{tab:specialvalues} presents special variables and values.
The value $\bullet$ identifies the $\textit{unknown}$ field of a given variable.
The $\textit{unknown}$ field of a variable is used to access statically-unknown indices of the variable.
In Tab.~\ref{projections}, there are several helper functions (projections) defined; we use them in the subsequent definitions.

\begin{table}
\begin{tabularx}{\columnwidth}{|cX|}
\hline
$s \in \Sigma$  &
= $Var \times Map \times Index \times Aliases$ \\

$m \in Map$  &
= $Var \rightarrow \PW{Val}$ \\

$i \in Index$  &
= ($Var \times Val) \rightarrow Var$ \\

$a \in Aliases$  &
= $Aliases_{must} \times Aliases_{may}$ \\

$a_{must} \in Aliases_{must}$  &
= $Var \times Var$ \\

$a_{may} \in Aliases_{may}$  &
= $Var \times Var$ \\
\hline

\end{tabularx}
\\
\caption{Data-flow analysis state-space.}
\label{tab:statespace}
\end{table}

\begin{table}
\vspace{-6pt}
\begin{tabularx}{\columnwidth}{|cX|}
\hline


$\textit{undefVar} \in Var$  &
is a variable representing an undefined variable. \\


$* \in Val$  &
is the statically-unknown value. \\

$\textit{undefined} \in Val$  &
is the undefined value. \\

$\bullet \in Val$  &
is the value representing index-name of $\textit{unknown}$ field. \\
\hline

\end{tabularx}
\\
\caption{Special variables and values.}
\label{tab:specialvalues}
\vspace{-5pt}
\end{table}

\vspace{-10pt}
\subsection{Data-flow equations}
\vspace{-3pt}
For propagating states through the nodes of the control-flow graph (CFG), we use a modification of standard data-flow equations for the forward problem.
Each node $k$ of CFG has six states associated:
$IN_k$, $GEN'_k$, $IN'_k$, $GEN_k$, $KILL_k$ and $OUT_k$. 
$IN_k$ represents the data coming to $k$; 
it is created by merging the states going out from all predecessors of $k$. 
In the case that the node has more predecessors, we call this state \emph{join point} and the operation \textit{mergeStates} merges information from different states.
If the node has only one predecessor, the operation \textit{mergeStates} only copies the information.
The state $GEN'_k$ defines variables that are newly defined by the update and data of these variables (note that these data can come only from unknown fields).
The state $KILL_k$ defines data that is removed from variables by the update while $GEN_k$ defines data that is added from the right-hand side of the update statement.
Finally, the state $OUT_k$ represents
updated data, i.e.,
the data going out from this node. 
The predicate $pred(k)$ returns the set of $n$ output states associated with the predecessors of $k$.

The data-flow equations are: 
\begin{align*}
IN_k &= \textit{mergeStates}(\{OUT_{p}\}), p \in pred(k)\\ 
IN'_k &= IN_k \cup GEN'_k\\
OUT_k &= GEN_k \cup (IN'_k - KILL_k)\\
\end{align*} 
\vspace{-3pt}
For the the initial node $i$, the output state is as follows: 
\begin{align*}
OUT_i  = (root, \{(root, \emptyset), (unk, \{\textit{undefined}\})\}, \{((unk, \bullet), root)\}, (\emptyset, \emptyset))
\end{align*} 
That is, the state contains variable $root$ representing a symbol table and a variable $unk$ which is its unknown field---it represents statically-unknown variables.

\begin{table}[ht]
\begin{tabularx}{\columnwidth}{|c|X|}
\hline
 \bfseries \#&   \bfseries Expression \DH{Zmenit nazvy mnozin, ..., aby zacinaly velkymi pismeny} \\
\hline
\hline

&  Let $x = (root, map, i, (a_{may}, a_{must}))$ be a state.\\

\hline

 \defitem&   $r(x) $\tabto{15em}$= root$ \\


 \defitem&   $values(x, v \in Var) $\tabto{15em}$= \{vals, (v \rightarrow vals) \in map\}$ \\



 \defitem&   $values(x, V \in \PW{Var}) $\tabto{15em}$= \{values(x, v), v \in V\}$ \\

 \defitem&   $values_{undef}(x, V \in \PW{Var}) $\tabto{15em}$= values(x, V)$\tabto{23em} if $V \not= \emptyset $ \\
 &   \tabto{15em}$= \{\textit{undefined}\}$\tabto{23em} if $V = \emptyset$ \\
 
  \defitem&   $values(x, \textit{undefVar})$\tabto{15em}$= \{\textit{undefined}\}$\DH{Mozna by tam tohle nemuselo byt, stacilo by pouzivat $values_{undef}$}\\



\hline

 \defitem&   $\textit{indexOf}(x) $\tabto{15em}$= i$ \\



 \defitem&   $indices(x, V \in \PW{Val}) $\tabto{15em}$= \{ iv, \exists_{v \in V} \exists_{n \in Val} ((iv, n), v) \in i \}$ \\

 \defitem&   $indices(x, V \in \PW{Var}, I \in \PW{Val}) $\tabto{15em}$= \{ iv, \exists_{v \in V} \exists_{ind \in I} ((iv, ind), v) \in i \}$ \\


\hline

 \defitem&   $aliases_{must / may}(x) $\tabto{15em}$= \{ (v_1, v_2), (v_1, v_2) \in a_{must / may} \vee (v_2, v_1) \in a_{must / may} \}$ \\



 \defitem&   $aliases_{must/may}(x, v \in Val) $\tabto{15em}$= \{a, (v, a) \in aliases_{must/may}(x)\}$ \\

 \defitem&   $aliases_{must/may}(x, V \in \PW{Val}) $\tabto{15em}$= \{ (a, v), a \in aliases_{must/may}(x, v), v \in V \}$ \\

 \defitem&   $aliases(x, v \in Val) $\tabto{15em}$= aliases_{may}(x, v) \cup aliases_{must}(x, v)$ \\

\hline

\end{tabularx}
\\
\caption{List of helper functions and projections.}
\label{projections}
\end{table}

%
%
%
%
%
%
%
%

\vspace{-10pt}
\subsection{Access Paths}


We describe expressions for accessing variables and associative arrays of an arbitrary depth using access paths.
An access path consists of a single value or a sequence of access paths:

\medskip

\begin{tabular}{ll}
$AP$ & $::= a, a \in Val$ \\

& $::= []([ AP ])^*$\\
\end{tabular}

\medskip


\noindent Each access path from the sequence represents the expression for accessing the level of an associative array given by the position of the access path in the sequence.
This makes it possible to perform accesses to any associative array of an arbitrary depth where at each level of the array the set of values used for indexing is specified with\DH{specified by?} 
an arbitrary read-access.
As we will see later, a read access using an access path returns a set of values, which can include the \textit{undefined} and the $*$ values. 

An access path describes an access from a given index (variable).
In the following, $[v]([AP])^*$ denotes an access path $[]([AP])^*$ from variable $v$.
Access paths express any PHP expression describing data-access without loss of information.
For example, consider the following PHP expressions and the corresponding access paths in the state $s$:
$\$a[\$b]$--$[r(s)][a][[r(s)][b]]$,
$\$\$a$--$[r(s)][[r(s)][a]]$, 
$\$a[\$b[\$c]][2]$--$[r(s)][a][[r(s)][b][[r(s)][c]]][2]$. 

\vspace{-50pt}
\subsection{Read Accesses}
\vspace{-10pt}

\begin{table}
\begin{tabularx}{\columnwidth}{|c|X|}
\hline
\bfseries \#& \bf Expression  \\
\hline
\hline

\defitem & $Eval(x, AP)$\tabto{6em} $= \{a\}$ \tabto{14em} \tabto{20em}if $AP = a, a \in Val$  \\

 & \tabto{6em}$= \{values(x, v), v \in Vars(x, AP) \}$ \tabto{20em} if $AP = [][AP]^*$ \\

\hline

\defitem & $Vars(x, AP)$ \tabto{6em}$= Vars(x, r(x), AP)$ \\

\hline

\defitem & $Vars(x, v \in Var, AP)$\tabto{9em} $= \{\textit{undefVar}\}$ \tabto{17em} if $AP = a, a \in Val$\tabto{37em}(a) \\
 
 & \tabto{9em}$= \{\textit{undefVar}\}$ \tabto{17em} if $AP = [][AP_1][AP_2]...[AP_n] \wedge Vars_n(x, v, AP) = \emptyset$\tabto{37em}(b) \\

 & \tabto{9em}$= Vars_0(x, v, AP)$ \tabto{17em} if $AP = []$\tabto{37em}(c) \\

 & \tabto{9em}$= Vars_n(x, v, AP)$ \tabto{17em} if $AP = [][AP_1][AP_2]...[AP_n] \wedge Vars_n(x, v, AP) \not= \emptyset$\tabto{37em}(d) \\

& \\

\defitem & $Vars_0(x, v, AP)$\tabto{7em}$= \{v\}$ \\

\defitem & $\forall_{i \in 1, ..., n}:$ \\


& \tabto{1em}$Vars_i(x, v, AP)$\tabto{7em}$= indices(x, Vars_{i-1}(x, v, AP))$ \tabto{28em} if $* \in Eval(x, AP_i)$ \tabto{37em} (a)\\
& \tabto{7em}$= indices_r(x, Vars_{i-1}(x, v, AP), Eval(x, AP_i))$ \tabto{28em} if $* \notin Eval(x, AP_i)$ \tabto{37em} (b)\\
 
\defitem & $indices_r(x, V \in \PW{Var}, I \in \PW{Val}) = $ $indices(x, V, I) \cup$ \tabto{37em} (a)\\
& \tabto{1em}$\bigcup _{v \in V, ind \in I} \{ u, ((u, \bullet), v) \in \textit{indexOf}(x) \wedge \not \exists_{w \in Vars}  ((w, ind), v) \in \textit{indexOf}(x) \}$  \tabto{37em} (b) \\

\hline
\end{tabularx}
\\
\caption{Definition of read accesses.}
\label{readaccesses2}
\end{table}

%
%
%
%
%
%
%
%

Tab.~\ref{readaccesses2} defines the read access.
$Eval$ defines the set of values accessible via an access path from a symbol-table variable at a given state (13).
$Vars$ defines the set of variables accessible via an access path from the symbol-table variable (14) or from an arbitrary variable (15).

If the access does not identify any variable, the set consists of the undefined variable $\textit{undefVar}$ (15-a)--(15-b).
Otherwise the set of variables is obtained by a traversal of the $\textit{indexOf}$ relation. 
The traversal starts from the specified variable (16).
At each level, the access can be performed either with a statically-unknown value (17-a) or with a set of statically-known values (17-b).
In the first case, the set of variables at the next level involves all indices of all variables at the current level.
Note that these indices include $\textit{unknown}$ fields.
If the access is performed with a set of statically-known values, the set of variables at the next level includes indices of all variables at the current level that are identified by the values (18-a).
Moreover, if the accessed index is not yet defined for a variable at the current level, the unknown field of the variable is added (18-b).
Note that it is not necessary to follow aliases.
The reason is that a write access copies the data to all possible targets, including all possible aliases.

If a set of values at each level of an access path contains exactly one value, the $Var$ function yields a single variable.
We use the notation $[var][v_1]..[v_n]$ in state $x$ to denote the variable $Vars(x, var, [][v_1]...[v_n])$.
If the state and the variable from which the access is performed is clear from the context, we write only $[][v_1]...[v_n]$.

\example{
Assume the read-access at line 5 in Fig~\ref{fig:phpcodeex1}.
The access is performed from the root variable of the state using the access path $[][arr][1]$.
The variable $[][arr]$ is defined at line 4, while the index $[][arr][1]$ of this variable is not defined.
Thus, the first level consists of variable $[][arr]$,
while the second level consists of its $\textit{unknown}$ field---$[][arr][\bullet]$.}

\vspace{-6pt}
\subsection{Write Accesses}
\vspace{-3pt}

Tab.~\ref{tab:updates_traversal} defines the $GEN'$ set, which contains variables that are created by the assignment and alias statements---the variables statically mentioned for the first time in the left-hand side of the statement\footnote{In PHP, variables are defined also when they are mentioned for the first time in the right-hand side of the alias statement, i.e., the statement \id{\$a = \&\$b} defines the variable \id{\$b} if it is not defined. While we model this behavior in our implementation, we ommit it from the formalization to make the presentation of our approach more clear.}.
It also defines the variables that are updated by these statements.
Tab.~\ref{tab:updates} defines $KILL$ and $GEN$ sets, which contains data that are removed and added to these variables.
Tab.~\ref{tab:deepcopy} defines the deep copy of a variable, which is used when a new variable is created and when an existing variable is assigned to another one.

\vspace{-7pt}
\subsubsection*{Collecting Variables}
\vspace{-4pt}

Tab.~\ref{tab:updates_traversal} defines four sets of variables. 
$Must$ and $may$ (22) are variables that either must or may, respectively, be updated by the assignment statement, $must'$ and $may'$ (23) are variables for the alias statement.
If a variable must be updated, a strong update
is performed---new information replaces current information.
If a variable only may be updated, a weak update is performed---new information is added to the information already present at the variable.

\DH{Vypustit tento odstavec?}
\example{ An example of a weak update is the update at line 4 in Fig.~\ref{fig:phpcodeex1}---it is not statically known which index of the variable $[][arr]$ is updated.
Weak updates are also performed, e.g., at line 19 in Fig.~\ref{fig:phpcodeex1}.
While variable $[][arr2][3]$ is strongly-updated, variables $[][arr][3]$ and $[][alias]$ that may be aliases of $[][arr2][3]$ are only weakly-updated.}

\begin{table}[ht]
\begin{tabularx}{\columnwidth}{|c|X|}
\hline\
 \bfseries \#&   \bfseries Expression \\
\hline
\hline

   \defitem &   $\textit{deepcopy}_{assign}(S \in \Sigma \times Var, T \in {\Sigma \times Var}):$ \\ 
        &  \tabto{1em}$(x_s, v_s) $\tabto{9em}$\in S \wedge$ \\
        & \tabto{1em} $(x_t, v_t) $\tabto{9em}$\in T \wedge$ \\
  &  \tabto{1em}$values(x_t, v_t) $\tabto{9em}$\supseteq values(x_s, v_s) \wedge$\tabto{25em} (a) \\
  &  \tabto{1em}$\forall_{v_{si} \in Var} \forall_{i_{si} \in Val} ((v_{si}, i_{si}), v_s) \in \textit{indexOf}(x_s) \implies ($ \tabto{25em} (b) \\
  &  \tabto{2em}$v_{ti} = \textit{createindex}(x_t, v_t, i_{si}) \wedge$ \\
  &  \tabto{2em}$\textit{deepcopy}((x_s, v_{si}), (x_t, v_{ti})) \wedge$  \\
  &  \tabto{2em}$(\not \exists_{v_{unkn} \in Var} ((v_{unkn}, \bullet), v_s) \in \textit{indexOf}(x) \wedge $ \\
  &  \tabto{3em}$v'_{unkn} $\tabto{10em}$= \textit{createindex}(x_t, v_t, \bullet) \wedge$ \\
  &  \tabto{3em}$values(x_t, v'_{unkn}) $\tabto{10em}$\supseteq \{\textit{undefined}\}))$ \\

  \defitem &   $\textit{deepcopy}(S \in \Sigma \times Var, T \in {\Sigma \times Var}):$\\
  & \tabto{1em} $ \textit{deepcopy}_{assign}(S, T) \wedge$ \\
     &  \tabto{1em}$(x_s, v_s) $\tabto{9em}$\in S \wedge$ \\
     & \tabto{1em} $(x_t, v_t) $\tabto{9em}$\in T \wedge$ \\
 &  \tabto{1em}$aliases_{must}(x_t, v_t) $\tabto{9em}$\supseteq \{a, a \in aliases_{must}(x_s, v_s)\} \wedge$ \\
  &  \tabto{1em}$aliases_{may}(x_t, v_t) $\tabto{9em}$\supseteq \{a, a \in aliases_{may}(x_s, v_s)\} \wedge$ \\
 
  \defitem &   $\textit{createindex}(x, parent \in Var, ind \in Val) = \{var, var = newvar(x) \wedge   $ \\
 &  \tabto{1em}$\textit{indexOf}(x)$\tabto{9em}$\supseteq \{((var, ind), parent)\} \wedge$ \\
  &  \tabto{1em}$values(x, var)$\tabto{9em}$\supseteq \{\textit{undefined}\} \wedge$ \\
 &  \tabto{1em}$aliases_{must}(x, var)$\tabto{9em}$\supseteq \{var\}\}$ \\

\hline

\end{tabularx}
\\
\caption{Definition of deep copy of the index.}
\label{tab:deepcopy}
\end{table}

\begin{table}
\begin{tabularx}{\columnwidth}{|c|X|}
\hline
\bfseries \#  &
\bfseries Expression \\
\hline
\hline

& \textit{Assignemnt / Alias:} \tabto{9em}$LHSAP = RHSAP / LHSAP = \&RHSAP$ \\


 & $LHSAP \sim [][AP_1][AP_2]...[AP_n]$ \\
 
 & $\forall_{j = 1, 2, ..., n} I_j = Eval(AP_j)$ \\

\hline

 \defitem & $must $ \tabto{27pt}$= must_n \wedge may $ \tabto{98pt}$= may_n$ \\
 
 \defitem & $must'$ \tabto{27pt}$= must'_n \wedge may' $ \tabto{98pt}$= may'_n$ \\


 \defitem & $indices_w(V \in \PW{Var}, I \in \PW{Val})$ \tabto{1em}$= indices(IN, V, I) \cup $ \\
 & \tabto{2em} $\bigcup _{v \in V, ind \in I} \{ \textit{defindex}(v, ind), \not \exists_{w \in Var} ((w, ind), v) \not\in \textit{indexOf}(IN \cup GEN') \}$ \\
 
 \defitem & $\textit{defindex}(v \in Var, ind \in Val)$ \tabto{1em}$=\{i,  ((u, \bullet), v) \in \textit{indexOf}(IN \cup GEN') \wedge$ \\
  & \tabto{2em} $i = \textit{createindex}(GEN', v, ind) \wedge$ \\
  & \tabto{2em}$\textit{deepcopy}((IN \cup GEN', u), (GEN', i))\}$ \\

\hline

\defitem & $must_0 =\{r(IN)\} \wedge may_0 = \emptyset$ \\

\defitem & $\forall_{j \in 1, 2, ..., n}$ \\
& \tabto{1em}$((|I_j| = 1 \wedge I_j \neq \{*\}) \wedge ($ \tabto{30em} (a) \\

 & \tabto{2em}$must_j $ \tabto{50pt}$= aliases_{must}(indices_w(must_{j-1}, I_j)) \wedge$\\

 & \tabto{2em}$may_j $ \tabto{50pt}$= aliases(IN'_k, indices_w(may_{j-1}, I_j)) \cup aliases_{may}(IN, indices_w(must_{j-1}, I_j)))) \vee$ \\

& \tabto{1em}$((|I_j| > 1 \wedge * \notin I_j) \wedge ($ \tabto{30em} (b) \\

 & \tabto{2em}$must_j $ \tabto{50pt}$= \emptyset \wedge$ \\

 & \tabto{2em}$may_j $ \tabto{50pt}$= aliases(IN'_k, indices_w(may_{j-1} \cup must_{j-1}, I_j)))) \vee$ \\

& \tabto{1em}$(* \in I_j \wedge ($ \tabto{30em} (c) \\

 & \tabto{2em}$must_j $ \tabto{50pt}$= \emptyset \wedge$ \\

 & \tabto{2em}$may_j $ \tabto{50pt}$= aliases(IN'_k, indices(IN'_k, may_{j-1} \cup must_{j-1}))) )$ \\

\hline

\defitem & $((I_n| = 1 \wedge I_n \neq \{*\}) \wedge ($ \\

 & \tabto{1em}$must'_n $ \tabto{40pt}$= indices^j_w(must_{j-1})) \wedge$ \\

 & \tabto{1em}$may'_n $ \tabto{40pt}$= indices^j_w(may_{j-1}))) \vee$ \\

 & $((|I_j| > 1 \wedge * \notin I_j) \wedge ($ \\

 & \tabto{1em}$must'_n $ \tabto{40pt}$= \emptyset \wedge$ \\

 & \tabto{1em}$may'_n $ \tabto{40pt}$= indices^j_w(may_{j-1} \cup must_{j-1}))) \vee$ \\

& $(* \in I_n \wedge ($ \\

 & \tabto{1em}$must'_n $ \tabto{40pt}$= \emptyset \wedge$\\

 & \tabto{1em}$may'_n $ \tabto{40pt}$= indices(IN'_k, may_{n-1} \cup must_{n-1})))$ \\

\hline

\end{tabularx}
\\
\caption{Definition of collecting variables for an update.}
\label{tab:updates_traversal}
\end{table}

Similarly to read accesses, the variables which are updated by a statement are defined by a traversal of the $\textit{indexOf}$ relation starting in the root variable of the state.
The traversal uses the access path of the left-hand side (\id{LHSAP}).
However, the traversal differs from that of a read access.
The first difference is that for a write access also the corresponding aliases are followed.
That is, all aliases whose data can be possibly changed are updated by the write access. 
That is why it is not necessary to follow the aliases during read accesses.
The second difference is that if a write access to an index identified by a statically known value is performed and this index does not exist, it is created. 

Creating new indices when traversing the $\textit{indexOf}$ relation is handled by the definition of the $indices_w$ set (24).
Note that write accesses to $\textit{unknown}$ fields in preceding program points could update also the newly defined index and its sub-indices.
Thus the new index contains a deep copy of the \id{unkwnown} field (19). 
That is, it contains all values and aliases from the $\textit{unknown}$ field (20)--(21-a) and also a deep copy of all indices of the $\textit{unknown}$ field (19-b).

\example{
The statement at line 12 in Fig.~\ref{fig:phpcodeex1} creates a new variable $[][arr][1]$.
The write access to the $\textit{unknown}$ field $[][arr][\bullet]$ at line 11 could update also this new variable and the data from this $\textit{unknown}$ field is thus copied to a new variable.
Thus the sub-index $[][arr][1][2]$ of the variable $[][arr][1]$ is defined.
\DH{?? Zkontrolovat: Will the update at line 10 add also undefined value?}}

The traversal begins with the $must_0$ set initialized with the variable $r(IN_k)$, which corresponds to the symbol table and the $may_0$ set initialized with the empty set (26).
The statement (27) describes a single step of the traversal at the level \id{j}.
The set $I_j$ of values used to access the $j$-th level of an associative array\DH{pouzivat associative array? Sjednotit.} can have (27-a) a single statically-known value, (27-b) several statically-known values, or (27-c) it can contain a statically-unknown value.


\example{
As an example of case (27-a), see the statement at line 9 in Fig.~\ref{fig:phpcodeex1}.
The $must_2$ set consists of  variable $[r(GEN')][arr][2]$, the $may_2$ set of variable $[r(IN)][alias]$, and the set of values $I_3$ consists of value $3$.
Thus, the $must_3$ set contains must aliases of the index $[r(GEN')][arr][2][3]$.
The only must-alias of this index is the index itself.
The $may_3$ set contains all aliases of  index $[r(GEN')][alias][3]$, which is again only the index itself and all may-aliases of  index $[r(GEN')][arr][2][3]$, which is the empty set.}

In (27-b) and (27-c), the $must_j$ set is empty---it is not known which index is accessed.
The $may_j$ set consists of all aliases of the indices of variables in $must_{j-1}$ and $may_{j-1}$.
In case of (27-c), the variables do not need to be identified by values of $I_j$ and no new variables are created.

\example{
As an example, see the statement at line 12 in Fig.~\ref{fig:phpcodeex1}.
At the second level, $must_2$ set consists of the variable $[r(GEN')][arr][1]$ and the $may_2$ set is empty.
The $must_3$ set is empty, while the $may_3$ set consists of variables $[r(GEN')][arr][1][2]$ and $[r(GEN')][arr][1][\bullet]$.}

The difference between computing variables that will be updated by the assignment and the alias statement (28) is caused  by the fact that the assignment statement updates the variable and all its aliases with new values while the alias statement un-aliases the variable from all its original aliases while keeping their  values unaffected.
However, the alias statement respects the aliases at previous levels the same way as the assignment statement. 
In other words, the expressions for obtaining indices to be updated for the assignment and the alias statements treat differently only the last level.

\example{
In the case of the alias statement at line 15 in Fig.~\ref{fig:phpcodeex1}, both variables $[r(GEN')][arr][2][1]$ and $[r(GEN')][alias][1]$ will be updated, since ($[r(GEN')][arr][2]$ is a may-alias of $[r(GEN')][alias]$).
\DH{Zkontrolovat, jestli $[][arr][2] neni take may alias [][arr][\bullet]$---to by nemel byt. Zkontrolovat, jak se to chova v kontextu prikazu na radku 19}
In the case of the alias statement at line 16, only the variable corresponding to $[r(IN)][arr][2]$ will be updated---the variable $r(IN)][arr]$ has no alias. 
\DH{Prikaz at line 19 je problematicky, mozna jeste probrat, co dela}}

\vspace{-8pt}
\subsubsection*{Performing Update}
\vspace{-4pt}

Tab.~\ref{tab:updates} defines how the collected variables are updated by the assignment and alias statements.
Expressions (29)--(31) describe the data that is removed from the variables.
Both the alias and assignment statements remove all the values and indices of all updated variables that were present in these variables before the update including the data added in collecting phase (29)--(30).
The alias statement also removes aliasing data (31).

Expressions (32)--(35) describe the data that is added to the variables.
In the case of a strong update (32), both the statements add just the data that results from merging variables obtained by the read accesses of the right-hand-side access path (RHSAP).
In the case of a weak update, the original variable is merged too, so the original data is preserved (33).
Technically, the update is described as first merging the data to a temporary fresh variable and then copying it from this variable to the variable being updated (34)--(35).
The difference between the assignment and the alias statements is that while the former one does not copy the alias data at the first level (34), the latter one does (35).
Note that for the other levels, the alias data are copied also in the case of the assignment statement (20-b).

\example{
As an example, see the update at line 20 in Fig~\ref{fig:phpcodeex1}.
The $must$ set is empty, the may set consists of variables $[][arr][1]$, $[][arr][2]$, $[][arr][\bullet]$, $[][arr2][1]$, $[][arr2][2]$, $[][arr2][3]$, and $[][arr2][\bullet]$.
Consider the update of variable $[][arr][1]$.
Because the update is weak, new data results from merging the result of read access of RHSAP, which is $[][arr2]$, with the variable that is updated, which is $[][arr][1]$.
Consequently, after the update, the variable $[]arr][1]$ contains indices $[][arr][1][1]$, $[][arr][1][2]$, and $[][arr][1][3]$.
E.g., index $[][arr][1][1]$ contains the data merged from indices $[]$$[arr]$ $[1]$$[1]$ and $[][arr2][1]$.}

\example{
Now consider the update at line 17.
The $must$ set consists of $[]$ $[arr2]$, the $may$ set is empty.
The read-access of the RHSAP results in reading $[][arr]$.
Because the update is strong, $[][arr]$ is the only variable that is merged and thus it is only deep-copied.
Note that in the case of the assignment statement, the alias data is copied for all the levels except for the first one.
Thus, because of the alias statement at line 16, $[][arr2][2]$ is a must-alias of $[][alias3]$ and $[][arr][2]$ and due to the alias statement at line 4, e.g., $[][arr2][\bullet]$ is a may-alias of $[][alias]$ and $[][arr2][\bullet]$.
Consequently, the statement at line 18 strongly updates not only $[][arr2][2]$, but also $[][arr][2]$ and $[][alias3]$.
Similarly, the statement at line 19 strongly updates $[][arr2][3]$ and weakly updates $[][arr][\bullet]$ and $[][alias]$.
Thus, the subsequent read access using access path $[][arr][3]$ would read also value $9$\DH{zkontrolovat}.}

\begin{table}
\begin{tabularx}{\columnwidth}{|c|X|}
\hline
\bfseries \#  &
\bfseries Expression \\
\hline
\hline

& \textit{Assignemnt / Alias:} \tabto{9em}$LHSAP = RHSAP / LHSAP = \&RHSAP$ \\



\hline

\defitem & $\forall_{v \in must \cup may \cup must' \cup may'}$\tabto{9em} $ values(KILL, v) $\tabto{16em}$= values(IN'_k, v)$ \\

\defitem & $\forall_{v \in must \cup may \cup must' \cup may'}$\tabto{9em} $ \textit{indexOf}(KILL, v) $\tabto{16em}$= \textit{indexOf}(IN'_k, v)$ \\

\defitem & $\forall_{v \in must'}$\tabto{9em} $ aliases(KILL, v) $\tabto{16em}$= aliases(IN'_k, v)$ \\



\hline


\defitem & $\forall_{v_{t} \in must \cup must'}$\tabto{6em} $ src = \{(IN, v), v \in Vars(IN, RHSAP)\}$ \\

\defitem & $\forall_{v_{t} \in may \cup may'} $\tabto{6em} $src = \{(IN, v), v \in Vars(IN, RHSAP)\} \cup \{(IN', v_t)\}$ \\

\defitem & $\forall_{v_{t} \in must \cup may}$\tabto{6em} $ \textit{deepcopy}_{assign}((GEN', v = \textit{fresh}(GEN')), (GEN, v_t)) \wedge mergeVars((GEN', v), src)$ \DH{parametry deepcopy nejsou spravne}\\

\defitem & $\forall_{v_{t} \in must' \cup may'}$\tabto{6em} $ \textit{deepcopy}((GEN', v = \textit{fresh}(GEN')), (GEN, v_t)) \wedge mergeVars((GEN', v), src)$\\

%
%


\hline

\end{tabularx}
\\
\caption{Definition of updates for assignment and alias statement and new object expression.}
\label{tab:updates}
\end{table}

\vspace{-8pt}
\subsection{Merge}
\vspace{-5pt}

Tab.~\ref{tab:merge} defines the merge operation.
Expression (36) defines the operation $\textit{mergeStates}$, which is used in the first data-flow equation to define the $IN$ state of a node.
It merges the root variables of the $OUT$ states of all predecessors of the node to the root variable in the IN state.
Note that if the node has only one predecessor, the merge actually corresponds to a deep copy.

\begin{table}
\begin{tabularx}{\columnwidth}{|c|X|}
\hline
\bfseries \#  &
\bfseries Expression \\
\hline
\hline

\defitem & $\textit{mergeStates}(OUT \in \PW{\Sigma}) = \{IN, mergeVars((IN, r(IN)), \{(o, r(o)), o \in OUT\})\}$ \\

\hline

\defitem & $mergeVars(R \in \Sigma \times Var, M \in \PW{\Sigma \times Var}):$ \\

& \tabto{1em}$APs = \bigcup_{(x, v_r) \in M} (accessPaths(x, v_r, [])) \cup \{[]\} \wedge$\tabto{28em} (a) \\

& \tabto{1em}$ResAPs = extend(APs) \wedge$\tabto{28em} (b) \\

& \tabto{1em}$\forall_{AP \in ResAPs} (mergeAP(R, M, AP))$\tabto{28em} (c) \\

\defitem & $mergeAP(R = (x_R \in \Sigma, v_R \in Var), M \in \PW{\Sigma \times Var}, AP):$ \\
& \tabto{1em}$resVar $\tabto{11em}$= \textit{createVar}(R, AP) \wedge$\tabto{37em}(a) \\

& \tabto{1em}$mergedVars $\tabto{11em}$= \bigcup_{(x_M, v_M) \in M} \{(x_M, Var(x_M, v_M, AP))\} \wedge$\tabto{37em}(b) \\


& \tabto{1em}$values(o, resVar) $\tabto{11em}$= \bigcup_{(x_m, v_m) \in mergedVars} \{values_{undef}(x_m, v_m)\} \wedge$\tabto{37em}(c)\\

& \tabto{1em}$aliases_{must}(x_R, resVar)$\tabto{11em}$= \bigcap_{(x_m, v_m) \in mergedVars} \{aliases_{must}(x_m, v_m)\} \wedge$\tabto{37em}(d) \\

& \tabto{1em}$aliases_{may}(x_R, resVar) $\tabto{11em}$= \bigcup_{(x_m, v_m) \in mergedVars} \{aliases_{must}(x_m, v_m) \cup aliases_{may}(x_m, v_m)\} -$\tabto{37em}(e) \\
& \tabto{2em}$ aliases_{must}(x_R, resVar) \wedge$ \\

& \tabto{1em}$indices(x_R, resVar) $\tabto{11em}$= \bigcup_{(x_m, v_m) \in mergedVars} \{\textit{createVar}((x_R, resVar), [][n]), $\tabto{37em}(f)\\
& \tabto{2em}$n \in indicesNames(x_m, v_m)\}$ \\


\hline

\defitem & $accessPaths(x \in \Sigma, v_R \in Var, AP) $\tabto{16em}$= \bigcup_{((i, i_{name}), v_R) \in \textit{indexOf}(x)} ($\\
& \tabto{1em}$ \{AP[i_{name}]\} \cup accessPaths(x, i, AP[i_{name}]) )$ \\


\hline

\defitem & $extend(APs) = APs \cup \bigcup_{l \in levels(APs) \wedge (AP \in APs \wedge level(AP) > l)} \{newAP(AP, l, values(APs, l))\}$ \\

\defitem & $newAP([][v_1]...[v_n], l \in Int, V \in \PW{Val}) $\tabto{180pt}$= \{[][u_1]...[u_n], \forall_{i=1, ..., n \wedge i \not= l} (u_i = v_i) \wedge u_{l} \in V\}$ \tabto{35em} if $v_{l} = \bullet$ \\
& \tabto{180pt}$= \emptyset$ \tabto{35em} if $v_{l} \not= \bullet$ \\

\defitem & $level([][v_1]..[v_n])$\tabto{180pt}$= n$ \\
\defitem & $levels(APs)$\tabto{180pt}$= \{l, l = level(AP) \wedge AP \in APs\}$ \\
\defitem & $value([][v_1]..[v_n], l \in Int)$\tabto{180pt}$= v_l$ \\
\defitem & $values(APs)$\tabto{180pt}$= \{v, v = value(AP) \wedge AP \in APs\}$ \\

\hline

\defitem & $indicesNames(x, var) = \{ n, \exists_{iv \in Var} \exists_{n \in Val} ((iv, n), var) \in \textit{indexOf}(x) \}$ \\

\hline

\end{tabularx}
\\
\caption{Definition of merge.}
\label{tab:merge}
\vspace{-5pt}
\end{table}

Expression (37) defines how variables in given states are merged into the resulting state.
Note that this operation is used also when an update is performed (34)--(35).
In (37-a), for each variable being merged and a state in which it is defined, the access paths of all sub-indices of the variable are collected.
The empty access path $[]$, which corresponds to the variables being merged, is added to these access paths (37-a). 

\example{
For merging at the join point at line 13 in Fig.~\ref{fig:phpcodeex1} and for the symbol-table variable of the first branch, the following access paths are collected: $[][alias]$, $[][alias][2]$, $[][alias][3]$, $[][arr]$, $[][arr][\bullet]$, $[][t]$, $[][t][2]$, $[][arr][1]$, $[][arr][1][2]$, $[][arr][1][3]$, $[][arr][2]$, $[][arr][2][3]$.}

Sub-expression (37-b) further extends the set of access paths.
After the extension, it contains an access path for each variable that will be defined in the resulting state.
For each access path that contains the value $\bullet$ (corresponding to the $\textit{unknown}$ field) at a certain level it adds the access paths that are created from this access path by replacing the value $\bullet$ with all the values that are in the input access paths at this level.
Note that this adds new access paths to the resulting set only if there were performed corresponding statically-known write accesses from different variables being merged.
This is analogous to copying indices of the $\textit{unknown}$ field when there is a write access 
with a given value for the first time and a new variable is thus defined.
While write-accesses to $\textit{unknown}$ fields in preceding program points could also create sub-indices of a newly defined variable, in the case of merge there could be write-accesses to $\textit{unknown}$ fields that could create sub-indices of 
variables created
elsewhere.
Both these operations are thus necessary to preserve the invariant that if there could be a write-access to an index using a statically known value at a given level, all the data that could be possibly written to this index are stored there.

\example{
When the merge at line 13 in Fig.~\ref{fig:phpcodeex1} is performed, the set of access paths is extended with $[][arr][2][2]$, which then causes the corresponding variable in the resulting state to be created.
The reason is that while in the else branch $[][arr][2]$ is not defined, there is a write access to the $\textit{unknown}$ field at line 11 that could create a sub-index of this variable.
The read access using $[][arr][2][2]$ will follow $[][arr][\bullet]$ in the second level and will finally $[][arr][\bullet][2]$, which contains values $6$ and \textit{undefined}.
In the then branch, $[][arr][2]$ is created and it will be thus added to resulting state of the merge.
The read access follows this variable at the second level.
Thus, to access value $6$ with access path $[][arr][2][2]$, there must be a variable $[][arr][2][2]$ which contains this value in the resulting state.
This is analogous to copying data from the $\textit{unknown}$ field to a variable that is statically stated for the first time when the update is performed, e.g, a new variable $[][arr][1][2]$ containing values $6$ and $\textit{undefined}$ is created during the update at line 12.}

Sub-expression (37-c) merges variables corresponding to an access path in the merged states to the variable which corresponds to this access path in the resulting state.
First the variable in the resulting state using the access path and the output variable is created (38-a).
Then, the corresponding variables in merged states are obtained (38-b).
Finally, the data of these variables are merged to the resulting variable (38-c)--(38-f).
Note that while in the case of (38-a), the access path is used to create the variable which directly corresponds to the access path, in the case of (38-b) the access path is used to get variables in merged states by the read access (15).
That is, in the case of (38-b), the variables can be accessible by $\textit{unknown}$ fields even at levels where the access path contains a static value.
Note that both (38-a) and (38-f) contain the expression $\textit{createVar}$, however, resulting variables are created only once---if expression $\textit{createVar}$ is used the second time with the same arguments, it returns the existing variable.

\example{
As an example, see the merge corresponding to the join point at line 13 in Fig.~\ref{fig:phpcodeex1}. 
For the access path $[][arr][1][3]$, variable $[][arr][1][3]$ is be created in the resulting state.
The merged variable in the first branch is $[][arr][1][3]$, however, for the second branch, the data corresponding to this access path is located in $[][arr][1][\bullet]$.
For the access path $[][arr][2][2]$, a variable is created in the resulting state, the merged variables are $\textit{undefVar}$ in the first branch and $[][arr][\bullet][2]$ in the second branch.}

\vspace{-6pt}
\subsection{Termination and Soundness}
\vspace{-3pt}
\paragraph{Termination:}
The values in our model are represented either by constants present in the program or values $*$, $\textit{undefined}$, and $\bullet$.
Thus the number of values is finite. 
From this it follows that the number of defined indices in a single dimension of arrays is finite.
However, due to the presence of loops and recursion, the infinite number of dimension may be generated.
To ensure the termination, the number of dimensions must be limited.
This can be done either explicitly~\cite{phantm} or implicitly by using, e.g, allocation-site abstraction for creating new dimensions of arrays.
Then, the total number of indices is finite and \textit{alias} and \textit{indexOf} relations are finite as well.
The transfer functions defined in Tab.~\ref{tab:updates_traversal}, Tab.~\ref{tab:updates}, and Tab.~\ref{tab:merge} are monotonic so the fixpoint computation terminates.

Our approach is implemented as a part of a static analyzer with the support of operators such as $+$.
Thus, potentially an infinite number of values can be generated in the program due to presence of loops and recursion.
To ensure termination of fixpoint computation in this case, it is necessary to limit the size of value sets of each variable by a constant---larger value sets would be represented either by value $*$ or by a finite abstract domain.

\vspace{-13pt}
\paragraph{Soundness:}
We use the following soundness argument: \begin{quote}If a value can be written to a given variable (index) by a write-access at the node $n_1$ of CFG, it is read from this variable by a read access in  node $n_2$ of CFG that follows $n_1$ if and only if there is a path from $n_1$ to $n_2$ in CFG where the variable is not strongly-updated by different value.
Moreover, if there is a path from the initial node to a given node such that a variable was not strongly-updated, the set of the variable values returned by a read access always includes the \textit{undefined} value.\end{quote}
Note that a value can be written to a given index also if the write access is statically unknown at any level.
Also note that there can be an arbitrary number of join points between $n_1$ and $n_2$.
We do not provide the proof of the argument, however, its validity follows from definitions of read accesses, write accesses, and merge. 


\vspace{-8pt}
\section{Implementation and Evaluation}
\vspace{-5pt}

We implemented our approach in the context of the Weverca \cite{weverca} static analyzer. Besides associative arrays, the implementation supports also objects.
Weverca makes it possible to perform static taint analysis~\cite{Tripp:2009:TET:1542476.1542486} as well as other analyses that can be used for finding bugs and evaluation the system safety.
It constructs CFGs on the fly.
If the processed program point is a virtual method call, it uses a read-access to get the method's receiver object and uses its type to determine the method to be called.
It also uses a read-access to get values of variables that specify the name of a method and the include target in case of a dynamic method call or dynamic include.
Weverca also implements several options of context-sensitivity for function and method calls.

The novelty of our approach is that it is sound and precise even if statically-unknown data from the input are used to access associative arrays-like structures at an arbitrary level.
Other approaches, such as ~\cite{Sridharan:2012} are more limited, e.g., they model only associative arrays of the depth 1.
Since it is not an issue to implement a precise analysis but a precise scalable analysis\DH{Tohle mi prijde moc shazujici cely pristup. Udelat sound a precise analyzu neni uplne trivialni. Pokusil bych se to rict nejak mirneji.}, a question we want to answer is whether our approach scales well.
In particular, we want to know how the merge function scales with respect to the number of variables being merged in the presence of dynamic write accesses to multiple levels of associative arrays.

To evaluate our approach we used the code \id{CODE\_n} that was generated from the code fragment at lines (2)--(19) in Fig.~\ref{fig:phpcodeex1} replicated $2^n$ times with all the variables except for the variable \id{\$any} with the prefix unique for each replica.
This code contains non-trivial dynamic accesses to multiple levels of associative arrays; the number of variables in this code grows exponentially with n.
However, the number of variables that are being merged at each join point is constant.
To evaluate the complexity of the merge operation, we use the code \id{mCODE\_n} (Fig~\ref{fig:phpcodeeval}) which is defined using \id{CODE\_n}.
In \id{mCODE\_n}, all variables defined in the code are involved in the top-level merge operation.

\begin{figure}[ht]

\begin{lstlisting}[language=PHP,numbers=none, xleftmargin=16em, xrightmargin=16em,frame=single]
$any = $_GET['user_input'];
if ($any) { CODE_n  }
else { CODE_n }
\end{lstlisting}
\caption{The code used for the evaluation}
\label{fig:phpcodeeval}
\vspace{-3pt}
\end{figure}

Tab.~\ref{table:evalresults} shows the results of analysis of \id{mCODE\_n} and \id{CODE\_n} for different values of $n$.
The table shows the number of nodes of generated CFG, the number of variables defined in our representation in the $OUT$ state of the program end point, the running time of the analysis.
The comparison of running times of \id{mCODE\_n} and \id{CODE\_n} shows that the merge operation is efficient even if all the variables all involved---\id{mCODE\_n} analyzes two copies of \id{CODE\_n} and additionally merges all the variables.
Thus the increase of time consumption is caused mostly by the increase of the number of variables.
This is caused mainly by the fact that the amount of data stored in a state grows with the number of variables defined in the state and that in the implementation there is no sharing of data between the states.
We believe that by optimizing the implementation, the scalability can be highly improved and the approach can scale up to thousands of variables and tens of thousands nodes of CFG.


\begin{table}
\centering
\begin{tabularx}{25em}{|c|X|X|X|}
\hline
\bfseries n &
\bfseries CFG nodes \newline \scriptsize \normalfont (mCODE\_n/CODE\_n)  &
\bfseries Variables &
\bfseries Analysis Time (s) \newline \scriptsize \normalfont (mCODE\_n/CODE\_n)
\\
\hline
\hline

 1&
235 / 117 &
107 &
0.4 / 0.3
 \\

 2&
463 / 231 &
211 &
1.3 / 0.7
 \\

 3&
919 / 459 &
419 &
4.9 / 2.5
 \\
 
  4&
 1831 / 915 &
 835 &
 22.8 / 10.3
  \\


\hline
\end{tabularx}
\\
\medskip
\caption{The evaluation results.}
\label{table:evalresults}
\vspace{-3pt}
\end{table}

\vspace{-12pt}
\section{Related work}
\vspace{-6pt}

In this section, we discuss tools and techniques related to the area of our
interest. 

Pixy~\cite{Jovanovic2006} is an open-source tool for detection of taint-style vulnerabilities in PHP 4. 
It involves a flow-sensitive, interprocedural,  and context-sensitive data flow analysis along with literal and alias analysis to achieve precise results.
The main limitations of Pixy include limited support for statically-unknown updates to associative arrays, ignoring classes and the \texttt{eval} command, and limited support for aliasing and handling file inclusion,
which all represent principle differences from programming languages such as
Java and C. 
Alias analysis introduced in Pixy incorrectly models aliasing between arrays and array indices. 
Web applications use associative arrays and objects extensively, thus we believe that this is an essential limitation.
Importantly, Pixy does not perform type inference, which also limits its precision and soundness.

Stranger~\cite{Yu2010} is an automata-based string analysis tool for PHP, which is built upon Pixy. It adds
a more precise string manipulation techniques that enable the tool to prove that an application is free from attack patterns specified as regular expressions.

Jang \cite{Jang:2009:PAJ:1529282.1529711} presents flow-insensitive points-to analysis for JavaScript. 
Their work is closely related to our approach, since JavaScript provides dynamic features similar to those in PHP.
The same way as our approach, they model variables, arrays, and objects using associative arrays.
However, they precisely model only assignments to constant indices and they use special unknown field for all other assignments.
Moreover, the same as Sridharan~\cite{Sridharan:2012}, they limit the accesses to associative arrays to depth one.
Next, PHP supports creating aliases between variables which JavaScript does not.
Therefore, we must maintain alias information to perform updates of points-to information correctly.

Phantm~\cite{phantm} is a PHP 5 static analyzer for type mismatch based on data-flow analysis;
it aims at detection of type errors. It combines run-time information from 
the bootstrapping phase of an application and static analysis when instrumentation using this information
is used. To obtain precise results, Phantm is flow-sensitive, i.e., it is able to handle situations 
when a single variable can be of different types. 
However, they omit updates of associative arrays and objects with statically-unknown values and aliasing, which can lead to both missing errors and reporting false positives.

F4F~\cite{Sridharan:2011} focuses on static taint analysis of web applications that use frameworks.
This work uses a semi-automatically generated specification of framework-related behaviors to reduce the amount of statically-unknown information, which arises, e.g., from reflective calls.

Andromeda static taint analyzer~\cite{Tripp:2013} fights the problem of scalability of taint analysis by computing data-flow propagations on demand.
It uses forward data-analysis to propagate tainted data and ignores propagation of other data.
If tainted data are propagated to the heap, it uses backward analysis to compute all targets to which the data should be propagated.
Andromeda analyzes Java, .NET, and JavaScript applications. 
The drawback of the approach is that it propagates only taint information.
The control-flow of the application can depend on other information which are then not available.

Sridharan~\cite{Sridharan:2012} et. al. present static flow-insensitive points-to analysis for JavaScript.
They model objects in JavaScript using associative arrays that can be accessed by arbitrary expressions.
However, they limit the accesses to depth one.
They show that in this setting, the complexity of flow-insensitive points-to analysis becomes $O(N^4)$, where $N$ is the program size, in contrast to the $O(N^3)$, which is the case when the accesses are constant.
To enhance the precision and scalability of the analysis, they identify correlations between dynamic property read and write accesses.
If the updated location and stored value can be accessed by the same first class entity (variable), it is extracted to a function parametrized by this entity; this function is then analyzed context-sensitively with the context be the variable.
Thus, the correlation between the update and store is preserved.

Schafer~\cite{Schafer:2013} et. al. present a dynamic analysis for identifying variables and expressions that always have the same value at a given program point and finding this value.
Such values can be used, e.g, to make a constrained dynamic language constructs static and thus enhance the scalability and precision of static analysis.

Wei~\cite{Wei:2013} et. al. reduce the number of statically-unknown information in static analysis by collecting such information at run-time.


Livshits~\cite{Livshits:2013} et. al. propose a method of fully automatic placement of security sanitizers and declassifiers.
They place sanitizers statically whenever possible and they try to minimize the amount of run-time tracking.
The input of their analysis is a data-flow graph generated by a static analyzer.
The quality of sanitization placement---the reduction of the amount of run-time tracking---depends on the quality of this data-flow graph.

\vspace{-7pt}
\section{Conclusion and future work}
\vspace{-3pt}
Dynamic languages such as PHP contain features that pose significant challenges for static analysis.
In our previous position paper~\cite{stpsa12} we introduced our approach to static analysis of PHP and described particular parts of the Weverca analyzer~\cite{weverca}.
In this paper we focused on the data modeling part.
We described dynamic accesses to associative array-like structures, which make it hard to apply value and points-to analysis,
and presented our approach to this challenge.
We focused on soundness and precision---we do not want to overwhelm the user with too many false alarms, which is often the case of related tools.

The evaluation shows that the prototype implementation of our approach scales to hundreds of variables and thousands of nodes in the control-flow graph.
We believe that the scalability of the tool can be further enhanced.
Therefore, as future work, we plan to enhance the scalability by implementing  optimizations, in particular, sharing the data between nodes of the control-flow graph, and then to evaluate the scalability on real-life applications.


\bibliographystyle{eptcs}
\bibliography{references}
\end{document}